\documentclass[aps,twocolumn,superscriptaddress,nofootinbib,a4paper]{revtex4-2}

\usepackage[T1]{fontenc}
\usepackage{times}
\usepackage{amsmath,amssymb,amsthm}
\usepackage{graphicx}
\usepackage{physics}
\usepackage{natbib}
\usepackage[colorlinks=true,linkcolor=blue,citecolor=magenta,urlcolor=blue]{hyperref}

\begin{document}
	
\title{Unbounded advantage of adaptivity in entanglement-assisted communication}

\author{Armin Tavakoli}
\affiliation{Physics Department and NanoLund, Lund University, Box 118, 22100 Lund, Sweden.}

\begin{abstract}
Shared entanglement can reduce the amount of classical communication needed to perform distributed computation. This can be achieved either by first measuring the entanglement and then communicating classically, or by adapting the quantum measurement to the received message.  Here, we introduce tasks that adaptive protocols can solve using just three entanglement-assisted messages, while the communication cost grows without bound when the quantum 	receiver is not adaptive. Consequently, no bounded amount of classical communication can reproduce all correlations generated by an entanglement-assisted ternary classical channel. We establish two complementary limitations on the relation  between adaptive and non-adaptive quantum protocols. First, every adaptive protocol using a finite entanglement dimension admits a non-adaptive simulation at finite communication cost. Second, every adaptive protocol sending $d$ messages can be simulated to within error $\varepsilon$ using $O(d\log(1/\varepsilon))$ non-adaptive messages. Our results reveal the fundamentally unbounded advantage of adaptivity in quantum communication, but also that  observing very large advantages requires both high-dimensional entanglement and very precise observations.
\end{abstract}

\date{\today}
\maketitle

\begin{figure}[t]
	\includegraphics[width=\columnwidth]{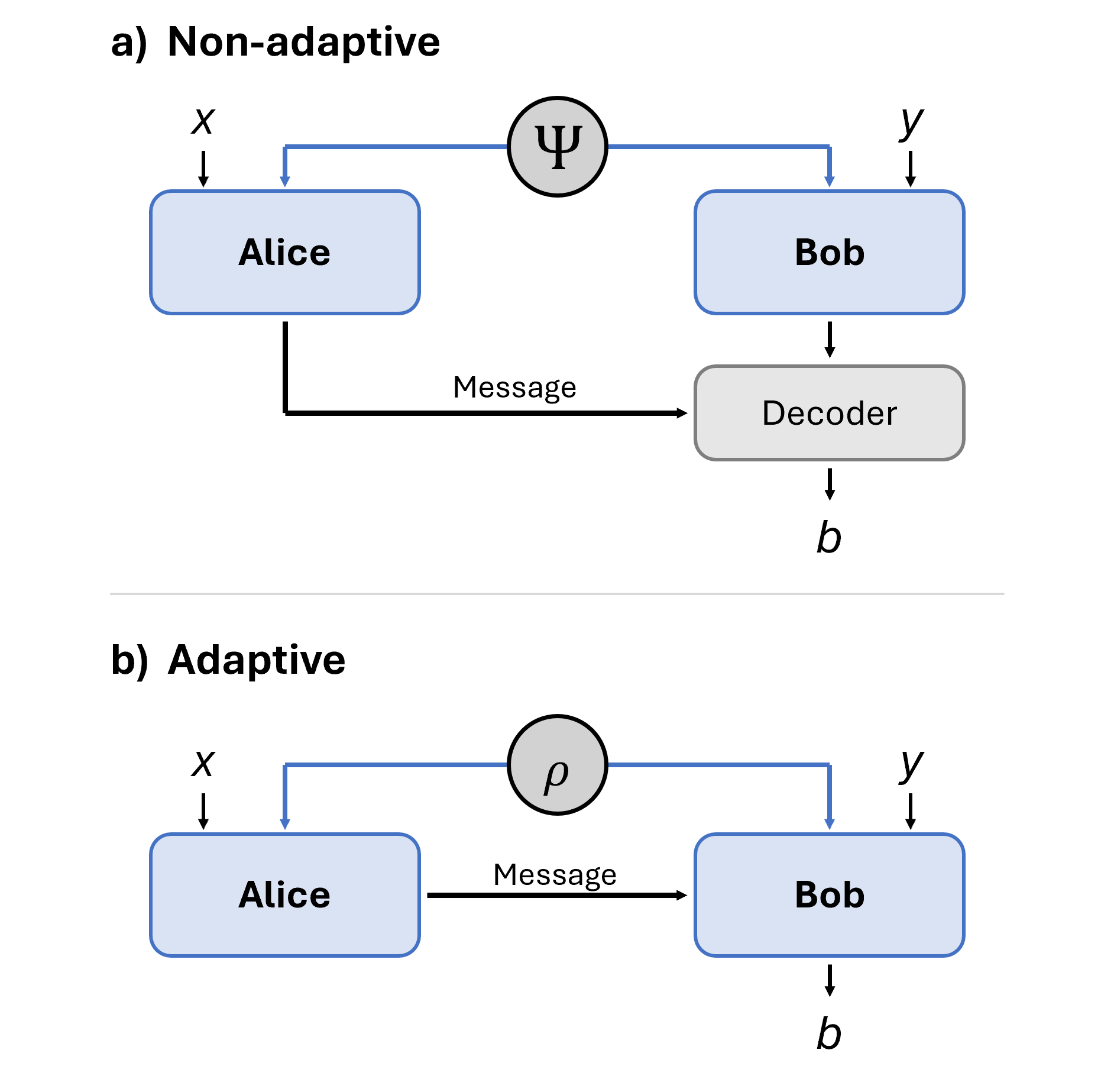}
	\caption{Entanglement is used to enhance classical communication. a) Non-adaptive protocols measure independently of the message and use it only to classically decode the quantum outcome. b) Adaptive protocols allow the quantum measurement to be  chosen after Alice's message is received.}
	\label{fig_scenario}
\end{figure}

\textit{Introduction.---} Shared entanglement allows separated parties to perform distributed tasks with less classical communication than is possible using shared randomness \cite{Buhrmann2010}. We study this type of quantum advantage in the prepare-and-measure scenario \cite{Brask2026, junior2026}, where Alice sends classical information to Bob, with the goal of computing a function whose inputs are distributed between them. The typical recipe is to first consume the entanglement by measuring the local shares in an appropriate Bell inequality test, and then use the strongly correlated local outcomes to enhance the power of classical communication \cite{Cleve1997, Brukner2004}; as illustrated in  Fig~\ref{fig_scenario}a. This has been exploited to reveal quantum advantage in a range of communication tasks (see e.g.~\cite{Pawlowski2009, Pawlowski2010, Pawlowski2012, Sadiq2014, Tavakoli2016, Hamedi2017, Pauwels2022b}). For exact communication tasks, such non-adaptive protocols can even yield exponential savings in the number of communicated bits compared with purely classical protocols \cite{Brassard1999}.

However, there is also a more general class of quantum protocols, where  Bob  does not need to measure before receiving Alice's message. He can instead use the message to choose which measurement to perform on his share of the entangled state; see Fig.~\ref{fig_scenario}b. Such adaptive protocols can sometimes outperform  their non-adaptive counterparts \cite{Pauwels2022}. Several recent works have investigated the role of adaptivity  for enhancing quantum advantage in a range of entanglement-assisted communication tasks \cite{Svegborn2026, Frenkel2022, Vieira2023, Alimuddin2023, Rout2025, Nogueira2026}.

In this Letter, we show that the advantage of adaptivity can be unbounded. Specifically, there exist tasks that an adaptive protocol solves with a ternary message, whereas every non-adaptive protocol requires a message alphabet growing without bound. The task is a simple exclusion game in which Bob has no input. In this case non-adaptive protocols reduce to classical communication with shared randomness, so no finite amount of classical communication can simulate an entanglement-assisted trit. This disproves the conjecture of Frenkel and Weiner \cite{Frenkel2022} that $d$ entanglement-assisted messages can always be simulated by $d^2$ classical ones. It also contrasts with their result that a bit assisted by maximal entanglement is worth no more than  two classical bits. This complements the recent finding that neither quantum communication nor Bell correlations of ququarts admit finite classical simulation \cite{deGois2026}; for us, the unbounded cost arises from entanglement assisting classical communication.

Our result raises the question of what it takes to obtain large separations, not only for our task but for any prepare-and-measure task, including those where Bob has an input. We answer from two directions. If Bob’s share of the entanglement has bounded dimension, the adaptive correlations can always be reproduced non-adaptively with a bounded message, so the unbounded advantage relies on entanglement dimension growing without bound. If we instead tolerate a small simulation error $\varepsilon$, Bob can simply guess Alice’s message in advance and let her report which guess succeeded. We show that the cost of this scheme is $O(d\log(1/\varepsilon))$, which diverges as $\varepsilon\to0$, but only slowly. The same idea yields exact simulations once any output noise is present. Witnessing a large adaptive advantage therefore requires an experiment that is near-ideal in both entanglement dimension and precision.

\textit{Adaptive and non-adaptive protocols.---} Alice and Bob  share a quantum state, as well as any amount of shared randomness. They receive independent inputs $x$ and $y$, respectively, and Alice sends a classical message to Bob, who produces an output $b$.  The amount of communication is quantified by the size of the message alphabet, i.e.~the number of possible values for the message.

In an adaptive protocol \cite{Pauwels2022}, Alice performs a $d$-outcome POVM $\{A_{a|x}\}_{a=1}^d$ on her part of the shared state $\rho_{AB}$ and sends its outcome $a$ to Bob. Conditioned on this message and his input, Bob performs a POVM $\{B_{b|y,a}\}_b$. The quantum correlations, generated by sending $d$-valued messages, become
\begin{equation}\label{eq_adaptive}
p(b|x,y)=\sum_{a=1}^d\tr\left(\rho_{AB}(A_{a|x}\otimes B_{b|y,a})\right).
\end{equation}
Any local processing can be absorbed into the respective POVMs, so this formula entails no loss of generality.

In a non-adaptive protocol, Bob's quantum measurement is independent of Alice's message. To clearly distinguish the adaptive and non-adaptive protocols, we use distinct notations. Bob performs a POVM $\{F_{c|y}\}_c$ with outcome $c$ on his share of the state $\Psi_{AB}$. After receiving a message $m\in\{1,\ldots,M\}$ from Alice, he produces $b$ according to some classical decoding distribution $r(b|m,c,y)$. This yields quantum correlations of the form
\begin{equation}\label{eq_nonadaptive}
q(b|x,y)=\sum_{m=1}^M\sum_c \tr\left(\Psi_{AB}(E_{m|x}\otimes F_{c|y})\right)r(b|m,c,y),
\end{equation}
where $\{E_{m|x}\}_m$ is Alice's POVM. Note that the cardinality of $c$ is unrestricted. In the special case when Bob has no input $y$, non-adaptive protocols are equivalent to classical communication assisted by shared randomness. Because without an input, the measurement correlations always admit a local hidden-variable model.

For an adaptive correlation $p(b|x,y)$ generated by Eq~\eqref{eq_adaptive}, its  non-adaptive simulation cost is the smallest alphabet size $M$ for which it can be reproduced by Eq~\eqref{eq_nonadaptive}, i.e~so that $p(b|x,y)=q(b|x,y)$. The non-adaptive simulation can use completely different states and measurements, and its shared state $\Psi_{AB}$ may have arbitrarily large dimension.

\textit{Unbounded adaptive advantage.---} We prove unbounded separation between adaptive and non-adaptive protocols by considering the following family of exclusion tasks indexed by an integer $M$. Consider the set of pairs $(j,a)$ such that $a\in\{1,2,3\}$ and $j\in\{1,\ldots,R\}$, where $R=M^2+1$. Alice's input is any subset, $S$, whose elements are pairs of this sort, such that $|S|=M$. Thus, the  set $S$ plays the role of $x$. Bob has no independent input. The goal is for Bob to output a pair of indices $b\notin S$ for every possible $S$ selected by Alice.

A simple argument shows that no non-adaptive protocol that sends $M$ messages can solve this task. Since Bob has no independent input, it suffices to consider classical protocols. Choose a specific value of the shared variable, i.e.~fix the deterministic encoding and decoding functions.  Let $\vec{b}=\{b_1,\ldots, b_M\}$ be Bob's list of outputs when receiving message $m\in\{1,\ldots,M\}$. Because Alice receives every subset $S$ of size $M$, one of them must contain the set $\vec{b}$. This means that there exists an input for which Bob is guaranteed to output $b\in S$ and hence fail the task. Since this holds for every encoding and decoding strategy, the failure persists for every classical protocol.

The less immediate argument is to prove that there exists an adaptive protocol that solves the task for any $M$ with just a ternary message ($d=3$). To this end, let the parties share the maximally entangled state $\ket{\phi^+_R}=\frac{1}{\sqrt{R}}\sum_{i=0}^{R-1}\ket{ii}$ and define $\{\ket{u_{j|a}}\}_{a,j}$ as a set of three mutually unbiased bases in dimension $R$. Such bases always exist \cite{Klappenecker2004}. Suppose that we could find a set of POVMs  $\{A_{a|S}\}_{a,S}$ with the property that  
\begin{equation}\label{lemma}
	A_{a|S}\ket{u_{j|a}}=0 \qquad \forall (j,a)\in S.
\end{equation}
This would immediately produce the desired protocol: Alice would measure her part of $\ket{\phi^+_R}$ using $\{A^T_{a|S}\}_{a,S}$, thereby remotely preparing the subnormalised state  $A_{a|S}/R$ for Bob. Then, Alice sends $a\in\{1,2,3\}$ to Bob who measures the corresponding basis $\{\ket{u_{j|a}}\}_{j}$. Eq~\eqref{lemma} guarantees that $b\notin S$.

Hence, the  main technical result is to prove that there exist POVMs satisfying Eq~\eqref{lemma}. To this end, for any given $S$ we construct matrices $V_a\in  \mathbb{C}^{R\times s_a}$ where the columns contain all the kets $\ket{u_{j|a}}$ with $(j,a)\in S$. The number of such kets is labeled $s_a$, so we have $s_1+s_2+s_3=M$. Write $T=[V_1, V_2, V_3]$ and consider the Gram matrix
\begin{equation}
G=T^\dagger T=
\begin{pmatrix}
\openone_{s_1} & G_{12} & G_{13}\\
G_{21} & \openone_{s_2} & G_{23}\\
G_{31} & G_{32} &\openone_{s_3}
\end{pmatrix},
\end{equation}
where $G_{kl}=V_k^\dagger V_l$. We now break the Gram matrix into three well-selected parts by writing $G=C_1+C_2+C_3$, where 
\begin{equation}
C_1=
\begin{pmatrix}
0 & 0 & 0\\
0 & \frac12 I_{s_2} & V_2^\dagger V_3\\
0 & V_3^\dagger V_2 & \frac12 I_{s_3}
\end{pmatrix}
\end{equation}
and analogously for $C_2$ and $C_3$; the $a$'th block-row and block-column are zero in $C_a$. These matrices are positive semidefinite whenever the off-diagonal block has operator-norm at most $1/2$. That is indeed always the case because for $k\neq l$
\begin{equation}
\|G_{kl}\|_\infty\leq \|G_{kl}\|_2 = \sqrt{\frac{s_k s_l}{R}}\leq \frac{M}{2\sqrt{R}}<\frac{1}{2},
\end{equation}
where the first equality uses the mutual unbiasedness and the last step uses $R=M^2+1$. Moreover, we have $\|G-\openone_M\|_\infty\leq\|G-\openone_M\|_2\leq\sqrt{M(M-1)/R}<1$ which means that the eigenvalues of $G$ cannot be zero and hence $G$ is invertible. We can now use this knowledge to construct our desired POVM. Let $P=TG^{-1}T^\dagger$ be the projector onto the span of the vectors associated with $S$ and define the POVM
\begin{equation}
A_{a}=TG^{-1}C_aG^{-1}T^\dagger+\frac{\openone-P}{3}.
\end{equation}
It is straightforward to check that $A_a\succeq 0$ and $\sum_a A_a =\openone$, so it is a valid POVM. To verify exclusion, take any vector  $\ket{u_{j|a}}=Te_k$ associated with $S$, where $e_k$ selects the column in $T$. Since this column belongs to block $a$ and the  $a$'th block-column of $C_a$ is zero, it follows that $C_aG^{-1}T^\dagger (Te_k)=C_aG^{-1}(T^\dagger T)e_k=C_ae_k=0$. Together with $(\openone-P)Te_k=0$, which holds by construction, this gives Eq~\eqref{lemma}. 

We have hence shown that for every $M$ an adaptive solution is possible using just three messages, whereas a non-adaptive solution is impossible using $M$ messages. Since this is true for every $M$, taking the limit  $M\rightarrow \infty$ reveals the unbounded advantage of adaptivity. Since Bob has no input, this is equally a separation between an entanglement-assisted trit and classical communication with shared randomness; in particular, the $d^2$ bound conjectured in \cite{Frenkel2022} fails already for $d=3$ .Next, we show that this unbounded advantage requires idealised conditions, for arbitrary prepare-and-measure tasks.

\textit{Finite entanglement dimension.---} Consider a general adaptive protocol using a $d$-valued message, where the dimension of Bob's share of the entangled state is at most $R$. We show that every such protocol admits an exact non-adaptive simulation whose communication cost is bounded solely by $d$ and $R$, independently of the numbers of inputs and outputs. Note that our above proof of the unbounded advantage of adaptivity indeed only arises in the limit $R\rightarrow \infty$.

Let us begin with the case where Bob has no independent input, as it is more simple and insightful. In an adaptive protocol, Alice's $d$-outcome POVM prepares for Bob the $R$-dimensional assemblage $\tau_{a|x}=\tr_A\left(A_{a|x}\otimes \openone \rho_{AB}\right)$. The correlations \eqref{eq_adaptive} can be written as
\begin{equation}
p(b|x)=\sum_{a=1}^d\tr(\tau_{a|x}B_{b|a})=\tr(\tilde{\tau}_x \tilde{B}_b).
\end{equation} 
In the second step,  we have defined the state $\tilde{\tau}_x=\bigoplus_{a}\tau_{a|x}$ and the POVM $\tilde{B}_b=\bigoplus_a B_{b|a}$. Both act on a Hilbert space of dimension  $D=dR$. Next, recall that non-adaptive protocols with no independent input on Bob are equivalent to classical protocols. Furthermore, the Frenkel-Weiner theorem \cite{Frenkel2015} states that correlations obtained from preparing and measuring $D$-dimensional quantum states without an input on Bob can be simulated classically with shared randomness using a $D$-dimensional message. Hence, every adaptive correlation can be simulated with a non-adaptive communication of size no more than 
\begin{equation}\label{eq_dimension_no_input}
M\leq dR.
\end{equation}
Thus, at fixed adaptive message size $d$, the simulation cost grows at most linearly with Bob's dimension $R$ and remains bounded for all tasks without an independent input on Bob.

A bound depending only on $d$ and $R$ also holds for general scenarios, where Bob has an independent input $y$.  Our approach again relies on the above block-encoding, namely that the adaptive correlations can be written as $p(b|x,y)=\tr\big(\tilde{\tau}_x \tilde{B}_{b|y}\big)$, where now  $\tilde{B}_{b|y}=\bigoplus_a B_{b|y,a}$. We choose to relax this set of correlations to the set of quantum correlations obtained in a prepare-and-measure scenario where Alice sends $D$-dimensional states to Bob,  and Bob performs arbitrary measurements on them. This simply means that we do not impose the block-structure and instead treat $\tilde{\tau}_x$ and $\tilde{B}_{b|y}$ as unconstrained $D$-dimensional states and measurements. Our proof, which is technical and detailed in supplementary material, shows that any such correlation can be simulated in a non-adaptive protocol with finite cost. The number of messages used in our simulation is far higher than \eqref{eq_dimension_no_input}, but it is most likely far from optimal.

\begin{figure}
	\includegraphics[width=\columnwidth]{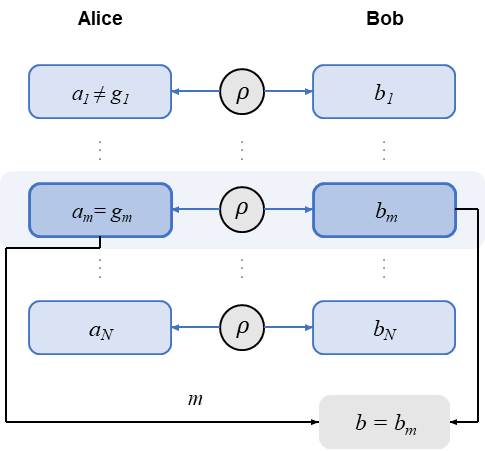}
	\caption{Bounded-error non-adaptive simulation of adaptive protocols. Bob guesses Alice's message through shared randomness and measures the POVM used in the adaptive protocol  on each of several independent copies of the shared state. Alice sends the index of the first correct guess, allowing  Bob  to select the corresponding output. Increasing the number of copies reduces the failure probability at the cost of a larger message alphabet.}\label{fig_simulation_epsilon}
\end{figure}

\textit{Simulation under bounded error.---} Let us now relax the previously fixed entanglement dimension and consider the non-adaptive simulation of arbitrary adaptive protocols, but we accept a bounded error, $\varepsilon>0$, in the simulation. We will quantify this error through the worst-case total-variation distance, defined as
\begin{equation}\label{eq_simulation_error}
\max_{x,y}\frac{1}{2}\sum_b |p(b|x,y)-q(b|x,y)|\leq\varepsilon.
\end{equation}
Here, $p$ and $q$ are the adaptive and non-adaptive correlations, respectively, from Eqs~\eqref{eq_adaptive} and \eqref{eq_nonadaptive}. The corresponding non-adaptive simulation cost of $p$, which we denote $M_\varepsilon$, is the smallest alphabet size so that this condition is satisfied. Below, we introduce a simple scheme that reveals key limitations on how small errors influence the simulation cost.  Its main  idea is illustrated in Fig~\ref{fig_simulation_epsilon}. Since Bob is not allowed to adapt his quantum measurement to Alice's message, our approach is to let him guess the message in advance and perform the corresponding POVM in the adaptive protocol. Then, we let Alice's message inform Bob of when his guess was correct. Bob can classically process this information to obtain an $\varepsilon$-accurate simulation.

To this end, consider that Alice and Bob share $N$ copies of the entangled state used in the adaptive protocol, namely $\rho_{AB}^{\otimes N}$. In addition, they share $N$ independent and uniformly random variables  $g_i\in\{1,\ldots,d\}$ for $i=1,\ldots, N$. These play the role of the guess. Specifically, for the $i$'th copy of the state, Bob treats $g_i$ as his guess for the message in the adaptive protocol. He therefore performs the adaptive-protocol measurement  $\{B_{b_i|y,g_i}\}_{b_i}$  on his share of the $i$'th copy and labels his outcome $b_i$. This is repeated independently for every $i$, leaving him with a string of outcomes $(b_1,\ldots,b_N)$. Alice  measures, independently for each copy of the state, the same POVM as in the adaptive protocol, namely $\{A_{a_i|x}\}_{a_i}$ for $i=1,\ldots,N$. She obtains the outcome string $(a_1,\ldots,a_N)$. For every $i$, Alice checks whether or not her outcome matches Bob's guess. In other words, she checks the condition $a_i=g_i$. The first instance at which the condition is fulfilled is marked by Alice as $m$. Thus, she sends the message $m=\min\{i:\hspace{1mm} a_i=g_i\}$ to Bob. When he receives it, he learns that his guess was correct for the $m$'th copy. Bob then discards all the other copies and outputs $b\equiv b_m$, with the guarantee that  $q(b|x,y)$ perfectly simulates the correlations in the adaptive protocol.

So far, Alice's non-adaptive simulation has used $N$ possible message values. However, there is also the possible event that none of Alice's $N$ measurements produce an outcome matching Bob's corresponding guess. In this event, Alice sends a distinct message, bringing the total cost to $M=N+1$ messages.  If Bob receives that 'all fail' message, he chooses to sample his output from some locally determined distribution $\tilde{q}(b|y)$. Since $g_i$ is uniformly random, the probability of a test failing is  $\delta=1-1/d$ independently of Alice's measurement. Thus, the probability of all $N$ tests failing is  $\delta^N$. The total-variation distance \eqref{eq_simulation_error} can therefore not exceed $\delta^N$, regardless of Bob's choice of $\tilde{q}$. This means that we must choose $N$ sufficiently large so that $\delta^N\leq \varepsilon$. Solving this for $N$ and correcting for the cost being an integer  gives
\begin{equation}\label{eq_approximate_cost}
M_\varepsilon\leq 1+\left\lceil\frac{\ln(1/\varepsilon)}{\ln(d)-\ln(d-1)}\right\rceil.
\end{equation}
Whether this  scaling is tight remains open.

This upper bound on the non-adaptive simulation cost holds for every adaptive protocol. Importantly, it diverges only logarithmically in the inverse error. It therefore provides the insight that every adaptive protocol  becomes expensive to simulate only when operating under very small   bounded error. For example, every ternary adaptive protocol can be simulated to total-variation error at most $10^{-6}$ using  $36$ non-adaptive message values, independently of its entanglement dimension or the number of inputs and outputs. We may also use that  $\ln(d)-\ln(d-1)\geq1/d$ to obtain the simpler formula  $M_\varepsilon \leq 1+\left\lceil d\ln(1/\varepsilon)\right\rceil$. For fixed $\varepsilon$, the upper bound in Eq~\eqref{eq_approximate_cost}, divided by the adaptive message size $d$, approaches $\ln(1/\varepsilon)$ as $d$ increases. Thus, the relative advantage  cannot rise faster than this slow-growth bound.

The approximate simulation scheme also provides an exact simulation when the adaptive correlations are subject to noise. Specifically, consider $p_t(b|x,y)=(1-t)p(b|x,y)+t\tilde{q}(b|y)$, where $t\in(0,1]$ is the noise fraction and $\tilde{q}$ is independent of Alice's input. This includes uniform output noise. Since the simulation produces $(1-\delta^N)p+\delta^N\tilde{q}$, choosing $\delta^N\leq t$ allows Bob to reproduce $p_t$ exactly (he needs to add further output noise when necessary to make up for the difference $t-\delta^N$). Thus, Eq~\eqref{eq_approximate_cost}, with $\varepsilon$ replaced by $t$, also bounds the exact non-adaptive simulation cost of these noisy adaptive protocols. Any fixed non-zero output-noise fraction therefore limits the advantage of any adaptive protocol.

\textit{Conclusion.---} We have shown that choosing a quantum measurement in response to a classical message can yield an unbounded communication advantage over both classical protocols and non-adaptive entanglement-assisted protocols. Our proof uses a family of tasks in which Bob has no independent input. A single trit suffices to solve every task adaptively, whereas the message alphabet required by non-adaptive protocols grows without bound. Thus, entanglement can give a fixed classical channel an unbounded advantage over classical protocols, and adaptivity can yield an equally unbounded advantage over non-adaptive protocols even when they have unrestricted shared entanglement. The former can equivalently be phrased as the signalling dimension \cite{DallArno2022} of an entanglement-assisted trit is infinite.

The advantage of adaptive over non-adaptive quantum protocols is, however, fragile. Finite entanglement dimension, any non-zero simulation error, or any input-independent output noise each renders the non-adaptive simulation cost finite. Very large adaptive advantages therefore require both high-dimensional entanglement and very precise observations.

Our bounded-error result suggests that unbounded simulation costs are tied to exactness. It would be interesting to determine whether the unbounded cost of classically simulating ququart communication \cite{deGois2026} similarly reduces to a slow divergence in $1/\varepsilon$ under approximate simulation. Another direction is adaptive communication in networks with more parties and sources \cite{Tavakoli2022, Bakhshinezhad2024, Lauand2025}. Their causal structure may obstruct the guessing-and-selection procedure used here, allowing the non-adaptive simulation cost to grow more rapidly as the tolerated error decreases. Identifying such networks could provide routes to large adaptive advantages under experimentally accessible conditions.

\begin{acknowledgements}
The author thanks Gabriele Cobucci, Elna Svegborn and Jef Pauwels for feedback on the manuscript. This work is financially supported by the Swedish Research Council under Contract No. 2023-03498, the Swedish Foundation for Strategic Research, and the Knut and Alice Wallenberg Foundation through the Wallenberg Center for Quantum Technology (WACQT).
	
\textit{AI statement:} All technical results in this manuscript were obtained from ChatGPT-6 Astra. The author has checked them for correctness and made them more presentable. Responsibility is assumed for the full content.
\end{acknowledgements}

\bibliography{references}

\appendix
\newpage
\section*{Supplementary Material}
\label{app_finite_simulation}

We prove that arbitrary state preparations and finite-outcome POVMs in dimension $D$ admit an exact non-adaptive simulation with a finite message alphabet depending only on $D$. Alice knows the classical description of the preparation, and Bob knows the POVM. Applying this result to the block encoding in the main text, with $D=dR$, establishes the claimed dimension-dependent simulation bound.

It suffices to consider a pure preparation $|\psi\rangle$ and rank-one POVMs $Q_b=\gamma_bP_b$, where $P_b=|\phi_b\rangle\langle\phi_b|$ and $\sum_b\gamma_bP_b=\openone_D$, so completeness implies $\sum_b\gamma_b=D$.  The  probabilities we aim to simulate are
\begin{equation}
	p_b=\gamma_bt_b,
	\qquad
	t_b=|\langle\psi|\phi_b\rangle|^2.
\end{equation}

\subsection{Protocol for arbitrary good approximations}
Our strategy is to first construct non-negative random scores for Bob, $w_b$, satisfying
\begin{equation}
	\mathbb E\!\left[w_b\right]=p_b,
	\qquad
	\sum_bw_b\leq B,
	\label{eq_em_scores}
\end{equation}
where $B$ depends only on $D$. If we had such scores, Bob could output $b$ with probability $w_b/B$ and otherwise reject, thus reproducing the target distribution conditional on acceptance. Since rejection has a probability of $1-1/B$, we could suppress it to any desired degree by repeating the procedure many times, at the cost of more communication. The protocol here will deliver this. However, this can never give a unit success probability, and that issue is addressed in the next subsection.

Our argument will rely heavily on the following lemma which we prove in subsection~\ref{SMlemma}. Let $n=D$, $L=2n+1$, and embed $\mathbb C^D$ in $\mathbb C^L$. There exists a bounded non-negative function $f$ on $[-1,1]$ such that, for every pair of unit vectors $\psi,\phi\in\mathbb C^D$,
\begin{equation}
	\mathbb E_{h,\theta}\!\left[
	\mathbf1\{|\langle\psi|h\rangle|^2>\theta\}
	f\!\left(\theta+|\langle\phi|h\rangle|^2-1\right)
	\right]
	=|\langle\psi|\phi\rangle|^{2n}.
	\label{eq_em_kernel}
\end{equation}
Here $h$ is a Haar-random ray in $\mathbb C^L$ and $\theta$ is independent and uniform on $[0,1]$. Moreover, $f(t)=0$ for $t\leq0$ and $f(t)\geq\kappa>0$ for $t>0$, with $\kappa$ depending only on $D$.

Equipped with this lemma, we now split the target probability into two  contributions,
\begin{equation}
	p_b=\gamma_b(t_b-t_b^n)+\gamma_bt_b^n.
	\label{eq_em_split}
\end{equation}
The first contribution is generated by a quantum experiment and the second will be supplied by Eq~\eqref{eq_em_kernel}. To this end, share $n$ copies of the maximally entangled state of local dimension $D$. Alice performs a binary measurement on all her copies, aiming to project onto  $(|\psi\rangle\langle\psi|^{\otimes n})^{\mathsf T}$ or else its complement. A successful projection is labeled $s=1$ (failed projection is $s=0$) and has probability $D^{-n}$. It prepares $\psi^{\otimes n}$ on Bob's $n$ registers. Without knowing $s$, Bob measures
\begin{align}\nonumber
	H_b&=\gamma_b\left(P_b\otimes I_D^{\otimes(n-1)}
	-P_b^{\otimes n}\right),\\
	H_*&=\sum_b\gamma_bP_b^{\otimes n}.
\end{align}
This measurement has possible outcomes $\{*\cup \{b\}\}$. Its meaning is that Bob measures $\{Q_b\}$ on the first register, obtains an outcome $b$ and then applies the POVM $\{P_b,\openone_D-P_b\}$ to each of the remaining $n-1$ registers. He records $*$ if all these $n-1$ tests give the first outcome, and otherwise leaves $b$ unchanged. We label by $z$ this (potentially) updated value and assign a score of $D^n$ if $z\neq *$ and $s=1$, and otherwise a score of zero. It follows that 
\begin{align}\nonumber\label{eqexpec1}
	\mathbb E_{s,z}&\!\left[D^ns\,\mathbf1\{z=b\}\right]\\
	&=D^n p(s=1) p(z=b| s=1)	=\gamma_b(t_b-t_b^n).
\end{align}
We have obtained the first part of \eqref{eq_em_split}.

For the second part, the parties use shared randomness to distribute $h,\theta$ as in Eq~\eqref{eq_em_kernel}. Alice computes the bit
\begin{equation}
	\chi=
	\begin{cases}
		1, & |\langle\psi|h\rangle|^2>\theta,\\
		0, & |\langle\psi|h\rangle|^2\leq\theta.
	\end{cases}
\end{equation}
After receiving $(s,\chi)$, Bob calculates the following score for every possible output label $b$,
\begin{align}\label{scores}
	w_b={}&D^ns\,\mathbf1\{z=b\}+\gamma_b\chi
	f\!\left(\theta+|\langle\phi_b|h\rangle|^2-1\right).
\end{align}

The roles of Alice's two bits are summarised in Table~\ref{tab_score_bits}. The bit $s$ determines whether the quantum contribution is retained, while $\chi$ determines whether the classical contribution is retained.

\begin{table}[h!]
	\centering
	\begin{tabular}{cc l}
		\hline
		$s$ & $\chi$ & Contributions to Bob's score $w_b$\\
		\hline
		$0$ & $0$ & Zero for every $b$\\
		$0$ & $1$ & Classical contribution only\\
		$1$ & $0$ & Quantum contribution only\\
		$1$ & $1$ & Sum of both contributions\\
		\hline
	\end{tabular}
	\caption{Dependence of the scores $w_b$ on Alice's two bits. A retained contribution may still vanish, depending on Bob's record or the argument of $f$. These are the scores before the correction used for deterministic exact simulation.}
	\label{tab_score_bits}
\end{table}

Combining Eqs~\eqref{eqexpec1} and \eqref{eq_em_kernel}, and averaging over both Alice's bits  and the shared randomness, gives
\begin{equation}
	\mathbb E_{s,z,h,\theta}\left[w_b\right]=\gamma_b(t_b-t_b^n)+\gamma_bt_b^n=\gamma_bt_b=p_b.
\end{equation}

When converting $\{w_b\}$ to probabilities, we pick up the acceptance probability of $1/B$. A simple bound on $B$ is 
\begin{equation}
	B=D^n+D\|f\|_\infty.
\end{equation}

\subsection{Providing an output when all trials reject}
Suppose the above protocol is repeated for $m$ rounds and all of them reject. Bob must still produce an output. We therefore introduce a distribution $q_b$ that he can sample in this event. Simply adding these outputs would generally change the target probabilities. To compensate, we identify an event on which a small contribution proportional to $q_b$ can be subtracted from the scores without making them negative.

Alice chooses an approximate description of her state, whose label she will include in her final message.  Choose finitely many reference states $v_1,\ldots,v_J$ such that every pure state is within distance $r=1/(16D)$ of at least one reference state, using trace-distance $d(\psi,v)=\sqrt{1-|\langle\psi|v\rangle|^2}$.  Alice sends a label $j$ satisfying $d(\psi,v_j)<r$.

Bob selects the measurement outcomes whose directions have sufficient overlap with $v_j$. For this, we define
\begin{equation}
	I_j=\{b:|\langle\phi_b|v_j\rangle|^2\geq1/(2D)\},
	\qquad
	\Gamma_j=\sum_{b\in I_j}\gamma_b.
\end{equation}
Measurement completeness gives
\begin{equation}
	1=\sum_b\gamma_b|\langle\phi_b|v_j\rangle|^2
	\leq\Gamma_j+\frac12.
\end{equation}
Here the selected outcomes contribute at most $\Gamma_j$, while the others contribute at most $(2D)^{-1}\sum_b\gamma_b=1/2$. Thus $\Gamma_j\geq1/2$. Bob can now sample from
\begin{equation}\label{qbdef}
	q_b=
	\begin{cases}
		\gamma_b/\Gamma_j,&b\in I_j,\\
		0,&b\notin I_j.
	\end{cases}
\end{equation}

We will later subtract a small contribution proportional to $q_b$ from the scores. To keep the corrected scores non-negative, we need an event on which $w_b$ is bounded below by a positive multiple of $q_b$ for every $b$. For the reference state $v_j$ specified by Alice, define the set of shared-randomness pairs
\begin{equation}
	S_j=\left\{
	(h,\theta):
	d(h,v_j)<r,\quad
	1-\frac1{4D}<\theta<1-\frac1{8D}
	\right\}.
\end{equation}
The event $S_j$ occurs when the sampled pair $(h,\theta)$ belongs to this set. These conditions ensure that $\chi=1$ and that
\begin{equation}
	\theta+|\langle\phi_b|h\rangle|^2-1>0
	\qquad\text{whenever }q_b>0,
\end{equation}
as we show below. The lower bound $f(x)\geq\kappa$ for $x>0$ will then give the required lower bound on the scores.

To calculate its probability, $\eta$, write $x=|\langle h|v_j\rangle|^2$. For a Haar-random ray in $\mathbb C^L$, $x$ has density $(L-1)(1-x)^{L-2}$, independently of $v_j$. Since $d(h,v_j)<r$ is equivalent to $x>1-r^2$, and $\theta$ is independent and uniform, we obtain
\begin{equation}
	\eta= \frac1{8D}\int_{1-r^2}^{1}(L-1)(1-x)^{L-2}\,dx
	=\frac{r^{2(L-1)}}{8D}>0.
\end{equation}
Here $1/(8D)$ is the length of the allowed threshold interval. On $S_j$, both $\psi$ and $h$ are close to $v_j$, so the triangle inequality gives $d(\psi,h)<2r$. Consequently,
\begin{equation}
	|\langle\psi|h\rangle|^2>1-4r^2>\theta,
\end{equation}
so Alice's bit is $\chi=1$. Moreover, replacing $v_j$ by the nearby vector $h$ changes any projector probability by less than $r$. Hence, for every selected outcome,
\begin{equation}
	|\langle\phi_b|h\rangle|^2
	>\frac1{2D}-r>\frac1{4D}.
\end{equation}
Together with $\theta>1-1/(4D)$, this makes the argument of $f$ positive. Since $f$ is then at least $\kappa$ (see the lemma), we obtain from the definitions \eqref{scores} and \eqref{qbdef} that
\begin{equation}\label{eq_em_domination}
	w_b\geq\kappa\gamma_b\mathbf1\{b\in I_j\}
	=\kappa\Gamma_jq_b
	\geq\frac{\kappa}{2}q_b
	\quad\text{on }S_j,
\end{equation}
where we used $\Gamma_j\geq 1/2$. Thus, on this event, every score contains a  contribution from $q_b$. This allows us to compensate in advance for the output produced when all rounds are rejected. Bob can identify the event using the shared randomness and Alice's reference-state label.

\subsection{Correcting the scores}
We now use this lower bound to correct the scores and complete the non-adaptive simulation protocol. The correction removes, in advance, the contribution that $q_b$ will supply when all $m$ rounds are rejected. Set $C=2B$ and choose a  number of rounds $m$ large enough such that
\begin{equation}\label{deltadef}
	\delta=(1-1/C)^m
	\leq\min\{1/2,\eta\kappa/2\}.
\end{equation}
The number $\delta$ will be the probability that all rounds reject. Since Bob then samples $q_b$, this event contributes $\delta q_b$ to the final output probabilities. We must therefore remove the same contribution from the other, accepted, rounds.

For each round, define the corrected scores
\begin{equation}
	\widetilde w_b=
	\frac{w_b-(\delta/\eta)\mathbf1\{S_j\}q_b}{1-\delta}.
	\label{eq_corrected_scores}
\end{equation}
Here, $\mathbf1\{S_j\}$ means that the subtraction is performed only when $S_j$ occurs. Since this event has probability $\eta$, the average amount subtracted is exactly $\delta q_b$. Moreover, the corrected scores remain non-negative. Outside $S_j$, nothing is subtracted; on $S_j$, Eq~\eqref{eq_em_domination} and the choice $\delta/\eta\leq\kappa/2$ guarantee that the amount subtracted does not exceed $w_b$.

At fixed preparation, POVM, and reference-state label $j$, we average over Alice's bit $s$, Bob's measurement record $z$, and the shared randomness $(h,\theta)$ of one round. The distribution $q_b$ is fixed under this average, while $\Pr[(h,\theta)\in S_j]=\eta$. Thus, Eq~\eqref{eq_corrected_scores} gives
\begin{equation}
	\mathbb E_{s,z,h,\theta}[\widetilde w_b]
	=\frac{\mathbb E_{s,z,h,\theta}[w_b]-(\delta/\eta)\eta q_b}{1-\delta}
	=\frac{p_b-\delta q_b}{1-\delta}.
	\label{eq_corrected_average}
\end{equation}
Summing Eq~\eqref{eq_corrected_average} over $b$ gives
\begin{equation}
	\mathbb E_{s,z,h,\theta}\!\Big[\sum_b\widetilde w_b\Big]
	=\frac{\sum_b p_b-\delta\sum_b q_b}{1-\delta}
	=\frac{1-\delta}{1-\delta}=1.
\end{equation}
Thus, the corrected scores sum to one on average (although their sum can vary between individual rounds). Their sum in any individual round is bounded by
\begin{equation}
	\sum_b\widetilde w_b
	\leq\frac{B}{1-\delta}
	\leq2B=C,
\end{equation}
where the last inequality follows from $\delta\leq1/2$, which holds by construction from Eq~\eqref{deltadef}.

These properties define a valid sampling procedure. Run $m$ independent rounds with the same preparation, POVM, and reference-state label $j$. After receiving Alice's message, Bob processes the recorded data in order. In each round, he outputs $b$ with probability $\widetilde w_b/C$ and rejects with the remaining probability $1-1/C\sum_b\widetilde w_b$.
He stops at the first accepted round. If every round rejects, he samples $q_b$. Bob outputs $b$ either in the first accepted round or, if all rounds reject, by sampling $q_b$. For round $i$ to be the first accepted round, the preceding $i-1$ rounds must reject, which occurs with probability $(1-1/C)^{i-1}$. The probability that round $i$ outputs $b$, averaged over the measurement outcomes and shared randomness of that round, is $\mathbb E[\widetilde w_b]/C$. Since the rounds are independent, these probabilities multiply. Summing over all possible accepted rounds and adding the contribution from total rejection gives
\begin{align}
	p_{\rm sim}(b)
	&=\sum_{i=1}^{m}\left(1-\frac{1}{C}\right)^{i-1}
	\frac{\mathbb E[\widetilde w_b]}{C}
	+\delta q_b\nonumber\\
	&=(1-\delta)\mathbb E[\widetilde w_b]+\delta q_b\nonumber\\
	&=p_b.
\end{align}
The second equality evaluates the geometric sum using $\delta=(1-1/C)^m$, and the last substitutes Eq~\eqref{eq_corrected_average}. Thus, the accepted rounds contribute $p_b-\delta q_b$, while the event that all rounds are rejected supplies the missing $\delta q_b$.

Note now the total communication cost of this non-adaptive simulation. Alice sends the two bits $(s,\chi)$ for in each of $m$ rounds, and also one reference-state label $j$. If there are $J$ reference states, this data can be seen as a single message with 
\begin{equation}
	F(D)=J4^m<\infty
\end{equation}
possible values. Both $J$ and $m$ depend only on $D$, which is finite by assumption since $D=dR$ where $R$ is the fixed  dimension of Bob's share of the entangled state. Hence, the simulation cost is finite. Since Alice’s message $(j, \{s_i,\chi_i\}_{i=1}^m)$ depends only on her preparation and the shared randomness, and $m$ depends only on $D$, the same message serves every POVM of Bob. The simulation therefore applies when Bob has an input $y$.

Finally, we have restricted to pure states and rank-one POVMs but this comes with no loss of generality. Mixed preparations are convex mixtures of pure ones, which Alice samples using shared randomness. Any POVM is a coarse-graining of a rank-one POVM, which Bob implements by classical post-processing of the simulated outcome.

\subsection{Proof of lemma}\label{SMlemma}

We prove the identity in Eq~\eqref{eq_em_kernel} by constructing the function $f$ explicitly. Fix the integer $n$ and put $L=2n+1$. We define
\begin{equation}
	u=|\langle\psi|h\rangle|^2,
	\qquad
	v=|\langle\phi|h\rangle|^2,
	\qquad
	a=|\langle\psi|\phi\rangle|.
\end{equation}
Here $\psi,\phi$ are fixed unit vectors in $\mathbb C^L$, $h$ is a Haar-random ray. We also let $\theta$ be independent and uniform on $[0,1]$. We seek a bounded non-negative function $f$ satisfying
\begin{equation}
	\mathbb E_{h,\theta}
	\left[\mathbf1\{u>\theta\}f(\theta+v-1)\right]
	=a^{2n},
	\label{eq_sm_goal}
\end{equation}
with $f(x)=0$ for $x\leq0$ and with a positive lower bound for $x>0$.

It is convenient to construct $f$ as the derivative of an auxiliary function. For this, we write
\begin{equation}\label{Gdef}
	G(x)=\sum_{k=1}^{L-1}c_k(x_+)^k,
\end{equation}
where $x_+=\max\{x,0\}$ and $c_k\geq 0$. Define
\begin{equation}
	f(x)=
	\begin{cases}
		G'(x),&x>0,\\
		0,&x\leq0.
	\end{cases}
\end{equation}
The indicator function in Eq~\eqref{eq_sm_goal} restricts $\theta$ to the interval $[0,u)$. Since $G(v-1)=0$, averaging over $\theta$ gives
\begin{align}
	\mathbb E_\theta
	\left[\mathbf1\{u>\theta\}f(\theta+v-1)\right]
	&=\int_0^u f(\theta+v-1)\,d\theta\nonumber\\
	&=G(u+v-1).
	\label{eq_sm_threshold}
\end{align}
We therefore need to choose the coefficients $c_k$ so that $\mathbb E_hG(u+v-1)=a^{2n}$.

To calculate this average, observe that
\begin{equation}
	u+v-1
	=
	\langle h|
	\left(|\psi\rangle\langle\psi|
	+|\phi\rangle\langle\phi|-\openone_L\right)
	|h\rangle.
\end{equation}
The operator in parentheses has eigenvalues $a,-a$, and  $-1$ (in $L-2$ copies). Let us use this eigenbasis $\{\ket{e_k}\}_k$ with eigenvalues $a$ and $-a$ associated to $\ket{e_1}$ and $\ket{e_2}$ respectively. Write $|h\rangle=\sum_{i=1}^L h_i|e_i\rangle$, with $\sum_i|h_i|^2=1$, and define
\begin{equation}
	z=|h_1|^2+|h_2|^2.
\end{equation}
Thus, $z$ is the probability that a measurement in this basis gives either of the first two outcomes. For a complex Haar-random vector, the weights $(|h_1|^2,\ldots,|h_L|^2)$ are uniformly distributed over the non-negative values summing to one. Their first two weights therefore have a sum with probability density
\begin{equation}\label{zdensity}
	p(z)=(L-1)(L-2)z(1-z)^{L-3},
	\qquad 0\leq z\leq1.
\end{equation}
The division of this weight between the two eigenvectors is uniform. We can therefore write their individual weights as $z(1+x)/2$ and $z(1-x)/2$, where $x$ is uniform on $[-1,1]$. It follows that
\begin{align}
	u+v-1
	&=a|h_1|^2-a|h_2|^2-\sum_{i=3}^{L}|h_i|^2\nonumber\\
	&=a\left[\frac{z(1+x)}{2}-\frac{z(1-x)}{2}\right]-(1-z)\nonumber\\
	&=azx+z-1.
\end{align}
Since $x\leq1$, the expression is non-positive for every $x$ whenever $z\leq1/(1+a)$. For larger $z$, it is positive precisely when $x>(1-z)/(az)$. Since $x$ is uniform on $[-1,1]$, averaging at fixed $z$ gives
\begin{align}
	\mathbb E_x[(azx+z-1)_+^k]
	&=\frac12\int_{(1-z)/(az)}^1\hspace{-7mm}(azx+z-1)^k\,dx\nonumber\\
	&=\frac{[z(1+a)-1]^{k+1}}{2az(k+1)}.
\end{align}
We then average over $z$ using its density \eqref{zdensity}. This yields
\begin{align}
	&\mathbb{E}_h\!\left[(u+v-1)^k_+ \right]
	=\frac{(L-1)(L-2)}{2a(k+1)}\nonumber\\
	&\qquad \times\int_{1/(1+a)}^1
	[z(1+a)-1]^{k+1}(1-z)^{L-3}\,dz.
\end{align}
Now make the variable substitution $q=(1-z)(1+a)/a$. The integral over $z$ evaluates to
\begin{align}
	&\int_{1/(1+a)}^1
	[z(1+a)-1]^{k+1}(1-z)^{L-3}\,dz\nonumber\\
	&\quad=
	\frac{a^{k+L-1}}{(1+a)^{L-2}}
	\int_0^1(1-q)^{k+1}q^{L-3}\,dq\nonumber\\
	&\quad=
	\frac{a^{k+L-1}}{(1+a)^{L-2}}
	\frac{(k+1)!(L-3)!}{(k+L-1)!},
\end{align}
where the last equality uses that the second-step integral is Euler's beta integral. Substituting this result into the expression for the Haar average gives
\begin{equation}
	\mathbb E_h\!\left[(u+v-1)_+^k\right]
	=
	\frac{k!(L-1)!}{2(k+L-1)!}
	\frac{a^{k+L-2}}{(1+a)^{L-2}}.
	\label{eq_sm_moment}
\end{equation}

We now choose the coefficients $c_k$ in \eqref{Gdef} to cancel the factorial factors in Eq~\eqref{eq_sm_moment}. Specifically, we choose
\begin{equation}
	c_k=
	2\binom{L-2}{k-1}
	\frac{(k+L-1)!}{k!(L-1)!},
	\qquad k=1,\ldots,L-1.
\end{equation}
This gives us
\begin{align}
	\mathbb E_h\!\left[G(u+v-1)\right]
	&=\sum_{k=1}^{L-1}c_k\,\mathbb E_h\!\left[(u+v-1)_+^k\right]\nonumber\\
	&=\sum_{k=1}^{L-1}
	\binom{L-2}{k-1}
	\frac{a^{k+L-2}}{(1+a)^{L-2}}\nonumber\\
	&=\frac{a^{L-1}}{(1+a)^{L-2}}
	\sum_{j=0}^{L-2}\binom{L-2}{j}a^j\nonumber\\
	&=a^{L-1}=a^{2n}.
\end{align}
The first equality follows from the definition of $G$ and linearity of expectation. The second uses the definition of $c_k$ and Eq~\eqref{eq_sm_moment}. The third uses $j=k-1$, and the last line follows from the binomial theorem and $L=2n+1$. Together with the threshold average in Eq~\eqref{eq_sm_threshold}, this proves Eq~\eqref{eq_sm_goal}.

All coefficients $c_k$ are positive. For $x>0$, the constant term in $f$ (obtained from differentiating the first term in $G$) therefore gives
\begin{equation}
	f(x)=c_1+\sum_{k=2}^{L-1}k c_kx^{k-1}
	\geq c_1=2L,
\end{equation}
so we may take $\kappa=2L$.  Since $f(x)=0$ for $x\leq0$ and $x^{k-1}\leq 1$, its maximum on $[-1,1]$ is
\begin{equation}
	\|f\|_\infty=f(1)=\sum_{k=1}^{L-1}k c_k<\infty.
\end{equation}
Both bounds depend only on $L=2n+1$, and hence only on $D$ when $n=D$. The positive lower bound permits the score subtraction, while the finite upper bound provides the normalisation constant $B$.
\end{document}